\documentclass{article}

\usepackage{arxiv}

\usepackage[utf8]{inputenc} 
\usepackage[T1]{fontenc}    
\usepackage{hyperref}       
\usepackage{url}            
\usepackage{booktabs}       
\usepackage{amsfonts}       
\usepackage{nicefrac}       
\usepackage{microtype}      
\usepackage{graphicx}
\usepackage{doi}
\usepackage{ulem}
\usepackage{graphicx}
\usepackage{geometry}
\usepackage{caption}
\usepackage{subcaption}
\usepackage{amsmath}
\usepackage{amssymb}
\usepackage{cuted}
\usepackage{wasysym}

\usepackage{ulem}
\usepackage[square,numbers]{natbib}

\usepackage{ulem}
\usepackage{graphicx}
\usepackage{geometry}
\usepackage{caption}
\usepackage{subcaption}
\usepackage{cuted}
\usepackage{wasysym}

\usepackage{ulem}

\title{Dynamics of Irreversible Particle Adsorption to Fluid Interfaces}

\author{  \hspace{1mm}Marina Pasquet$^{1,2}$\textsuperscript{*},
\hspace{1mm}Yu Fu$^1$\textsuperscript{*},
\hspace{1mm}Joelle Frechette$^{1}$.\\
\\
	$^1$ Chemical and Biomolecular Engineering Department, University of California,
Berkeley, \\
Berkeley, 94760, California, United States.\\
	$^2$ Biofisika Institute (CSIC, UPV/EHU) and Department of Biochemistry and Molecular Biology, \\
    University of the
Basque Country, Leioa, 48940, Spain. \\
\\
\textsuperscript{*} Marina Pasquet and Yu Fu contributed equally to this work.
}

\renewcommand{\shorttitle}{\textit{Dynamics of Irreversible Particle Adsorption to Fluid Interfaces}}

\hypersetup{
pdftitle={Dynamics of Irreversible Particle Adsorption to Fluid Interfaces},
pdfauthor={Marina Pasquet, Yu Fu, Joelle Frechette},
pdfkeywords={Particle adsorption , Interfacial dynamics , Random Sequential Adsorption (RSA) , Thiele modulus , Kinetically-limited adsorption , Pickering emulsions},
}

\begin{document}
\maketitle

\begin{abstract}
Understanding the dynamic adsorption of colloidal particles at fluid interfaces is essential for applications ranging from emulsion stabilization to interfacial assembly of functional materials. Adsorption dynamics is often described through diffusion-limited models (such as the Ward-Tordai framework) along with assuming dynamic equilibrium between the adsorbed and dispersed particles. However, most experiments show that particle adsorption is irreversible, and diffusion-limited models fail as the surface coverage goes beyond the dilute limit where particle crowding limits further adsorption. Here, we present a unified model that captures the transition from diffusion-limited to kinetic-limited regimes by coupling diffusion with a Random Sequential Adsorption (RSA)-based boundary condition that accounts for irreversible adsorption and particle blocking for a spherical droplet. Using both a microtensiometer and pendant drop tensiometry, we measure dynamic interfacial tension changes for 3-(Trimethoxysilyl)propyl methacrylate (TPM) particles at the toluene/water interface across a range of bulk concentrations, drop sizes, and particle functionalization. Our analysis shows that the adsorption flux becomes increasingly hindered as the surface area fills, in agreement with RSA predictions. Furthermore, we calculate the Thiele modulus as a dimensionless number that quantifies the relative importance of adsorption kinetics to diffusion. We find that above a critical surface coverage, adsorption becomes reaction-limited, marking a transition to kinetically controlled dynamics. This approach provides a predictive framework for particle adsorption at fluid interfaces and highlights the necessity of moving beyond equilibrium diffusion-limited models.
\end{abstract}

\section{Introduction}
The precise arrangement of colloidal particles at fluid-fluid interfaces represents a powerful paradigm for creating materials by design~\cite{hill2017a,cates2008a,aussillous2001a,thompson2015a,chen2014a,smirnov2015a,shipway2000a}. This approach has enabled the fabrication of a new generation of materials, including bijels, functional capsules, and even reconfigurable liquid circuits~\cite{fink2024afm}, where the collective behavior of the interfacial particles dictates the macroscopic properties and rheology of the interface itself~\cite{fu2024a,vialetto2020a,vialetto2021a,peito2022a,shin2019a,park2014a,vethaak2021a,pan2023a}.

Compared to traditional molecular surfactants, the appeal of colloidal particles lies in their substantial adsorption energies—often thousands of $k_B T$—which lead to effectively irreversible attachment and the formation of exceptionally robust, solid-like interfacial assemblies~\cite{peito2022a,shin2019a,park2014a,du2010a,binks2002a,berg2009a}. Recent studies have also demonstrated the versatility of particle-stabilized systems in applications ranging from cosmetics and pharmaceuticals to advanced functional materials~\cite{peito2022a,park2014a,shin2019a}.

Nevertheless, the fundamental processes controlling the adsorption of colloidal particles remain poorly understood. Particle adsorption extends beyond simple diffusion-driven transport from bulk solution to the interface, encompassing irreversible attachment processes as well as dynamic interfacial phenomena such as particle crowding, structural ordering, and jamming transitions~\cite{cui2019sa, chai2020,guzman2022a,du2010a,hua2018a,cui2017a,kaz2012a,garbin2012a}. The complexity of these interfacial dynamics renders purely diffusion-based models inadequate for accurately describing the particle assembly process. 

Classical theoretical frameworks for surfactant adsorption dynamics, notably the Ward–Tordai model~\cite{ward1946a,prosser2001a,slavchov2017a}, cannot be applied directly to particle adsorption because these models assume reversible attachment characterized by a dynamic equilibrium at the interface. However, in the case of particles, the attachment is effectively irreversible, meaning desorption events are negligible unless the particles are very small or undergo change in their surface energy, therefore no true equilibrium can be established~\cite{berg2009a,binks2002a,bizmark2014a,adamczyk2000a}. More critically, as the interface becomes densely populated, adsorption kinetics become increasingly dominated by crowding phenomena. Interfacial crowding significantly slows the adsorption rate by reducing the accessible interface area for new particles, often described as a "blocking" effect~\cite{bizmark2014a,adamczyk2000a}. Additionally, particle mobility at crowded interfaces is substantially reduced, manifesting as a coverage-dependent surface diffusivity that prevents the effective rearrangements needed to incorporate new particles~\cite{schwenke2014a,wang2016a}.

Experimentally, disentangling intrinsic interfacial phenomena from diffusion-limited mass transport remains extremely challenging. This challenge motivated the development of advanced experimental techniques, such as the microtensiometer, to minimize diffusion timescales and enable direct probing of intrinsic interfacial adsorption kinetics~\cite{alvarez2010a,alvarez2010b}. However, diffusion typically remains the limiting factor even at the small length scales of microtensiometry, underscoring the profound difficulty of isolating true kinetic limitations in particle adsorption experiments~\cite{hua2018a,hua2016a,manga2016a}.

In this work, we directly address these challenges by developing a unified experimental and theoretical framework for studying the adsorption dynamics of colloidal particles. We aim to capture the complete adsorption process, from initial diffusion-dominated kinetics to later stages governed by interfacial crowding and blocking. We utilize two distinct experimental platforms: microtensiometry, employing small toluene droplets with radii $r_d \approx 80  \mu$m, and pendant drop tensiometry, using larger droplets ($r_d \approx 1$ mm). Our experimental results are interpreted using a comprehensive theoretical approach coupling bulk diffusion with a Random Sequential Adsorption (RSA) kinetic model at the interface. The RSA model explicitly accounts for irreversible adsorption and coverage-dependent blocking effects characteristic of particulate systems~\cite{talbot2000a,adamczyk2005a,feder1980}.

This article is structured as follows: we first describe our theoretical framework, contrasting classical diffusion-limited models with our unified diffusion-RSA approach. We then detail the experimental materials and methods. In the Results and Discussion section, we validate our model against experimental observations, determine key kinetic parameters, and use the Thiele modulus to analyze the transition from diffusion-controlled to kinetically controlled adsorption regimes. Finally, we summarize our main findings and outline directions for future research.

\section{Theoretical Framework}

\begin{figure}
\centering
\includegraphics[width=1\linewidth]{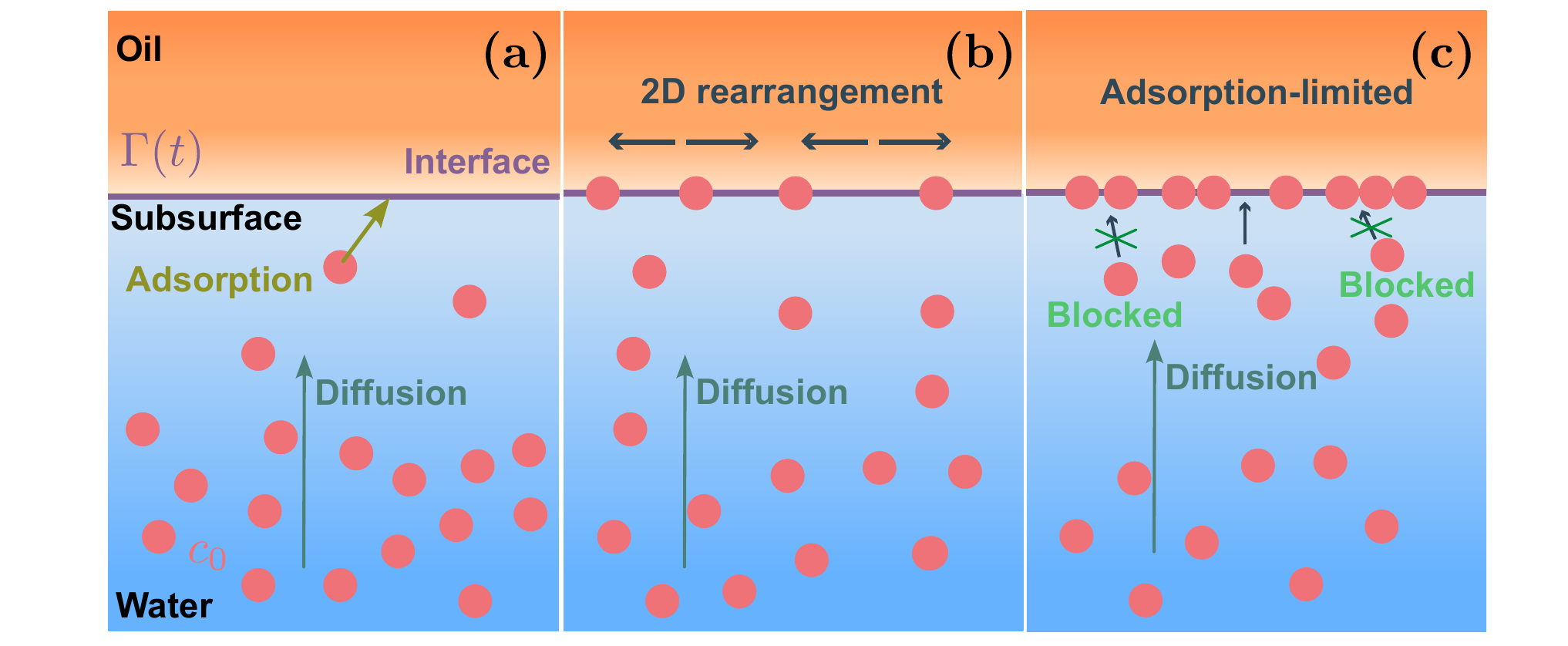} \caption{(a) Particle adsorption occurs in two primary steps: diffusion from the bulk water phase to the subsurface, followed by adsorption to the interface. (b) At low coverage, particles at the interface may undergo 2D-reorientation. (c) In the kinetically-limited regime, particles already adsorbed at the interface create steric hindrance, blocking further adsorption of incoming particles.}
\label{fig:schema_intro}
\end{figure}

 The adsorption of colloidal particles from a bulk dispersion onto a fluid interface can be conceptualized as a multi-step process (Fig. \ref{fig:schema_intro}). Initially, particles must be transported from the bulk phase to a region immediately adjacent to the interface, often termed the subsurface layer. This transport is typically governed by diffusion. Once a particle reaches the subsurface, it must overcome any potential energy barriers to successfully attach to the interface. For particles with high adsorption energy, this attachment step is irreversible.

Following attachment, particles may undergo further processes at the interface, such as 2D diffusion (Fig. \ref{fig:schema_intro}b), although mobility is often significantly hindered at higher surface coverages $\theta$. If transport is slow compared to attachment, the process is diffusion-limited. Conversely, if attachment is the slower step, often due to energy barriers or limited available space from steric blocking (Fig. \ref{fig:schema_intro}c), the process is kinetically-limited.

\subsection{Diffusion-Controlled Models}
\label{Part_WTTheo}
The Ward–Tordai model provides an analytical description for diffusion-controlled adsorption to a planar interface \cite{ward1946a}. It relates the time-dependent surface excess concentration, $\Gamma(t)$, to the history of the subsurface concentration, $c_s(t)$, via an integral equation:
\begin{equation}
\Gamma(t) = 2 c_0 \sqrt{\frac{D t}{\pi}} - \sqrt{\frac{D}{\pi}} \int_0^t \frac{c_s(\tau)}{\sqrt{t - \tau}} d\tau
\label{eq:WT_integral_cs_planar_detailed}
\end{equation}
Here, $c_0$ is the constant bulk concentration far from the interface, and $D$ is the particle diffusion coefficient.  To solve this, an additional relationship between $\Gamma(t)$ and $c_s(t)$ is required,  typically an adsorption isotherm assuming local dynamic equilibrium. For the Ward-Tordai model used in this work for comparison, we follow the approach for particles developed by Hua et al.~\cite{hua2018a}, which adapts the classical framework for molecular surfactants~\cite{lin1990a}. In this modified model, the partitioning between the subsurface concentration $c_s(t)$ and the surface excess $\Gamma(t)$ is still governed by an adsorption isotherm (e.g., Frumkin), which presumes a local equilibrium. However, the resulting change in interfacial tension is not calculated from the isotherm's corresponding equation of state. Instead, for irreversibly adsorbed particles where the adsorption energy ($\Delta E$) is much larger than thermal energy, the surface pressure, $\Pi(t) = \gamma_0 - \gamma(t)$, can be directly related to the surface concentration, $\Gamma(t)$, through a wetting equation of state~\cite{du2010a}:
\begin{equation}\label{eq:wetting_eos}
\Pi(t) = \Delta E \cdot \Gamma(t) = \gamma_0 \cdot \eta(t)
\end{equation}
where $\eta(t)$ is the surface coverage. This relationship allows the experimentally measured interfacial tension to be directly converted to surface coverage.

For adsorption onto curved interfaces, such as the droplets used in our experiments, mass transport is governed by Fick's second law in spherical coordinates:
\begin{equation}
\frac{\partial c(r,t)}{\partial t} = D \frac{1}{r^2} \frac{\partial}{\partial r} \left( r^2 \frac{\partial c(r,t)}{\partial r} \right)
\label{eq:fick_spherical_detailed}
\end{equation}
where $c(r,t)$ is the concentration at radial position $r$ and time $t$. The curvature introduces a geometric focusing effect, altering diffusion pathways compared to planar interfaces, particularly when the droplet radius $r_d$ is small \cite{alvarez2010a, alvarez2010b}.

\subsection{Irreversible Adsorption with Steric Blocking (RSA Model)}
\label{Part_RSATheo}
For colloidal particles, the adsorption energy is frequently very large compared to thermal energy, making the process effectively irreversible. Furthermore, particles occupy a significant area, and sterically hinder the adsorption of subsequent particles. This excluded area effect becomes the dominant factor limiting adsorption as the surface coverage increases.

To model this irreversible process, we modify the kinetic boundary condition at the interface to incorporate a blocking function, $B(\theta)$, derived from the principles of Random Sequential Adsorption (RSA) \cite{feder1980, talbot2000a}. The kinetic rate equation for irreversible adsorption becomes:
\begin{equation} \label{eq:adsorption_rate_rsa_model}
\frac{d\Gamma(t)}{dt} = k_a \ c(r_d,t) \ B\left(\frac{\Gamma(t)}{\Gamma_m}\right) = k_\text{a,eff}(\theta) \ c(r_d,t)
\end{equation}

\begin{figure}[htbp]
  \centering
  \includegraphics[width=0.7\linewidth]{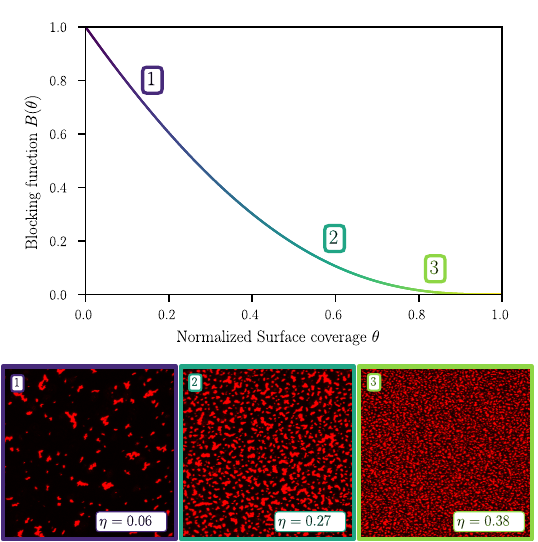}
  \caption{Blocking function and representative images of the interface at selected coverages.
    (Top) The RSA blocking function $B(\theta)$ decreases sharply as normalized surface coverage $\theta = \eta / \eta_m$ increases. The numbered boxes (1–3) correspond to the images below.
    (Bottom) Confocal fluorescence images of TPM-F108 particles (diameter 270~nm) at the toluene/water interface acquired at (1) $\eta=0.13$, (2) $\eta=0.58$, and (3) $\eta=0.81$, illustrating how crowding increasingly hinders further adsorption. Each image is $62\,\mu\text{m} \times 62\,\mu\text{m}$.}
  \label{fig:blocking_with_images}
\end{figure}

Here, $k_a$ is the intrinsic adsorption rate constant, $c(r_d,t)$ is the subsurface concentration at the interface, and $\Gamma_m$ is the maximum surface coverage at the RSA jamming limit. The function $B(\theta)$ represents the blocking function, where $\theta = \Gamma(t)/\Gamma_m$ is the normalized surface coverage. The effective adsorption rate constant is defined as:
\begin{equation}
k_\text{a,eff}(\theta) = k_a \cdot B(\theta)
\end{equation}
This definition captures how crowding at the interface dynamically reduces the available area for adsorption. As $\theta$ increases, $B(\theta)$ decreases from $B(0) = 1$ to zero, reflecting the transition from a clean to a fully jammed interface. Consequently, $k_\text{a,eff}$ declines over time, slowing the adsorption rate as the surface coverage increases.
For the blocking function, we use the well-established empirical form for hard spheres/disks \cite{adamczyk2000a}:
\begin{equation}\label{eq:blocking_function_rsa_model}
B(\theta) = \left( 1 + 0.812 \theta + 0.426 \theta^2 + 0.072 \theta^3 \right) (1 - \theta)^3 \quad \quad \quad 
\end{equation}
The highly non-linear nature of $B(\theta)$ reflects the rapidly increasing difficulty of finding a sufficiently large empty space for a new particle as the interface becomes crowded (Fig. \ref{fig:blocking_with_images}).

\subsection{Unified Diffusion-RSA Model}

A comprehensive model must account for both bulk diffusion and RSA kinetics. This is achieved by coupling the diffusion equation (Eq. \ref{eq:fick_spherical_detailed}) with a boundary condition at the interface ($r=r_d$) that equates the diffusive flux to the kinetically limited adsorption rate (Eq. \ref{eq:adsorption_rate_rsa_model}):
\begin{equation}
- D \frac{\partial c}{\partial r}\bigg|_{r = r_d} = k_a c(r_d,t) B\left(\frac{\Gamma(t)}{\Gamma_m}\right)
\label{eq:bc_rsa_model}
\end{equation}
This unified model (Eqs. \ref{eq:fick_spherical_detailed} \& \ref{eq:bc_rsa_model}) addresses the key mechanisms of irreversible particle adsorption onto curved interfaces: bulk diffusion, geometric effects of curvature, and kinetic limitation due to steric blocking described by RSA.

\section{Materials and Experimental Methods}

\subsection{Materials}

The following reagents were used as received: 3-(Trimethoxysilyl)propyl methacrylate (TPM, 98\%, Sigma-Aldrich), 2-hydroxy-4$^\prime$-(2-hydroxyethoxy)-2-methylpropiophenone (photoinitiator, 98\%, Sigma-Aldrich), sodium chloride (NaCl, Sigma-Aldrich), sodium hydroxide (NaOH, Fisher Scientific), poly(ethylene glycol)-block-poly(propylene glycol)-block-poly(ethylene glycol) (Pluronic F108, Sigma-Aldrich), Tween 20 (Fisher Scientific), toluene ($>$ 99.9\%, Fisher Scientific), hydrogen peroxide, 200 proof KOPTEC ethanol (VWR), sulfuric acid (95\%, J.T. Baker), and RBS 35 solution concentrate (Sigma-Aldrich). Xiameter OFS-6124 silane (Dow Corning) was used for surface modification. Deionized (DI) water with resistivity greater than 18.2 M$\Omega\cdot$cm was obtained from a Milli-Q Integral Water Purification System.

\subsection{Particle Synthesis and Characterization}

TPM colloidal particles were synthesized using a spontaneous emulsification method followed by photopolymerization, based on procedures detailed elsewhere \cite{neibloom2020a}. Briefly, an aqueous solution (10 mL) was prepared containing NaOH and NaCl to achieve a pH of 11.9 and an ionic strength of 15 mM, along with a stabilizing surfactant (either Pluronic F108 or Tween 20). Liquid TPM monomer (110 $\mu$L), containing 0.2\% photoinitiator, was carefully added onto the aqueous phase. The molar ratio ($\omega$) of TPM to surfactant was controlled: $\omega = 50$ for Pluronic F108 and $\omega = 45$ for Tween 20. The mixture was left undisturbed for 96 hours to allow spontaneous emulsification.

Subsequently, the TPM droplets were solidified into particles via UV-initiated radical polymerization. The resulting solid particles were purified by at least eight cycles of centrifugation, decanting, and redispersing in fresh DI water. Particle diameter and polydispersity were characterized using dynamic light scattering (Zetasizer Nano-ZS, Malvern Instruments). The average diameter was 86 nm for TPM-F108 particles and 60 nm for TPM-Tween particles. Both systems had a Polydispersity Index (PDI) below 0.1.

\subsection{Interfacial Tension Measurements}

\begin{figure}[htbp]
\centering
\includegraphics[width=1\linewidth]{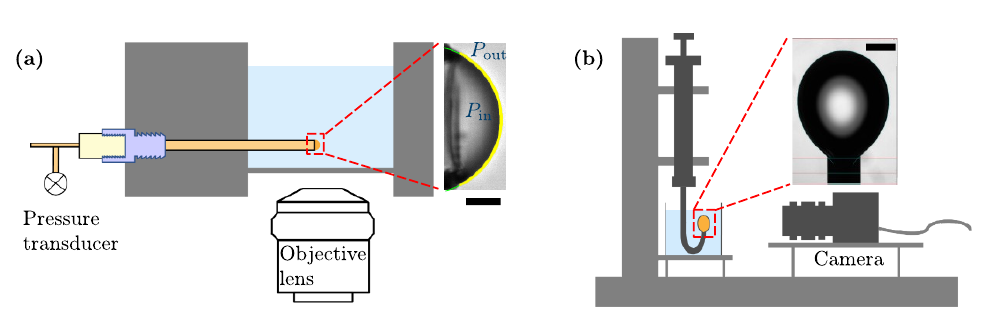}
\caption{(a) In the microtensiometer, a toluene drop (radius $r_d$) is held at a capillary tip. The pressure difference $\Delta P$ is measured, and the IFT is calculated via the Young-Laplace equation. The green line shows a circular fit to the drop contour. Scale bar: 20 $\mu$m. (b) In the pendant drop tensiometer, the IFT is calculated from the shape of a larger toluene drop. Scale bar: 1 mm.}
\label{fig:dispo_exp}
\end{figure}

Dynamic interfacial tension (IFT) was measured using two complementary techniques:

\begin{itemize}
    \item \textbf{Microtensiometry:} This technique utilizes small droplets ($r_d \approx 80 \mu$m). A schematic is shown in Fig. \ref{fig:dispo_exp}A. Additional details on the setup are available in this article \cite{alvarez2010a}. A hydrophobized glass capillary was filled with toluene and immersed in the aqueous particle dispersion. A toluene droplet was formed and held at a constant size by a controlled pressure head ($\Delta P$). The dynamic IFT was calculated from the Young-Laplace equation: $\gamma(t) = \Delta P(t) r_d(t) / 2$, where $\Delta P(t)$ and the droplet radius $r_d(t)$ were continuously monitored.

    \item \textbf{Pendant Drop Tensiometry:} This technique employs larger droplets ($r_d \approx 1$ mm) (Fig. \ref{fig:dispo_exp}B). A toluene droplet was formed from a needle in the aqueous dispersion. The dynamic IFT, $\gamma(t)$, was determined by fitting the full drop shape to the Young-Laplace equation, accounting for gravity.
\end{itemize}
Rigorous cleaning procedures were implemented for all experiments \cite{iasella2022a}.

\subsection{Numerical Methods}

The unified Diffusion-RSA model, consisting of the bulk transport equation (Eq.~\ref{eq:fick_spherical_detailed}) coupled with the non-linear interfacial rate equation (Eq.~\ref{eq:bc_rsa_model}), does not admit a closed-form analytical solution. The primary difficulty lies in the non-linear coupling between the evolving bulk concentration profile and the interfacial coverage, governed by the RSA blocking function $B(\theta)$. To solve this system, we developed a custom numerical model in \texttt{Python}, employing a finite difference scheme similar to those used in related surfactant and particle adsorption problems \cite{adamczyk2000a}.

The bulk diffusion process was modeled by solving Fick's second law in spherical coordinates. The semi-infinite radial domain was truncated to a finite computational domain, $r \in [r_d, L]$, with the far-field boundary $L$ set sufficiently large so as not to influence the interfacial dynamics over the simulation timescale. This spatial domain was discretized on a uniform grid. The Crank--Nicolson method was used to discretize the diffusion equation. This implicit, second-order accurate method is well-suited for diffusion problems. 

The simulation proceeds with a time-marching algorithm. At each time step:

\begin{enumerate}
    \item The bulk concentration profile is updated using the Crank--Nicolson scheme, incorporating appropriate boundary conditions. At the far-field boundary ($r = L$), a Dirichlet condition is applied: $c(L, t) = c_0$. At the droplet interface ($r = r_d$), the adsorption kinetics (Eq.~\ref{eq:bc_rsa_model}) are implemented as a Robin-type boundary condition. This condition dynamically links the diffusive flux from the bulk to the available surface area for adsorption.
    
    \item The solution of the diffusion equation provides the updated subsurface concentration $c(r_d, t)$.
    
    \item This subsurface concentration is then used to integrate the differential equation for the surface coverage $\Gamma(t)$ (Eq.~\ref{eq:adsorption_rate_rsa_model}). This integration is performed using a high-order, adaptive step-size solver: the Runge--Kutta--Fehlberg method (RK45), as implemented in \texttt{SciPy}.
\end{enumerate}

This numerical approach accurately captures the dynamic interplay between bulk transport and the increasingly hindered adsorption at the interface. It enables robust comparisons with experimental data obtained from both pendant drop and microtensiometer measurements.

\section{Results and Discussion}

\subsection{Dynamic Adsorption and Validation of the Diffusion-RSA Framework}

The dynamic adsorption of colloidal particles at the toluene/water interface was quantified by monitoring the time evolution of the interfacial tension (IFT, $\gamma$) over a range of bulk particle concentrations ($c_b$). We converted the measured surface pressure, $\Pi(t) = \gamma_0 - \gamma(t)$, into the surface coverage, $\eta(t)$, using the wetting equation of state (Eq.~\ref{eq:wetting_eos}). For Pluronic F108-stabilized particles, the maximum area fraction was experimentally determined as $\eta_{\text{max}} \approx 0.65$, while Tween 20-stabilized particles reached a lower limit of $\eta_{\text{max}} \approx 0.45$. These values were used to normalize the coverage data accordingly. We define the normalized coverage relative to the RSA jamming limit as $\theta(t) = \eta(t)/\eta_{\text{max}}$, where $\eta_{\text{max}}$ is the maximum fractional surface coverage. This allows for a unified comparison across systems with different particle chemistries and packing constraints.

\begin{figure}[tbhp]
\makebox[\linewidth][c]{%
  \includegraphics[width=1\linewidth]{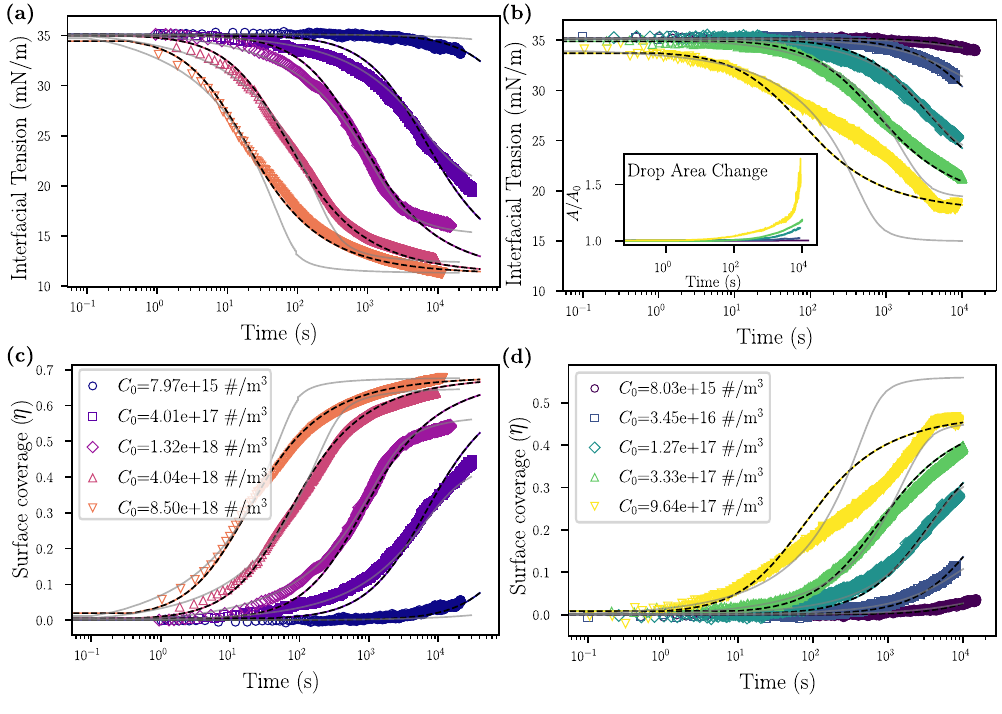}%
}
\captionsetup{width=1\linewidth}
\caption{Time‐evolution of (a,b) interfacial tension ($\gamma$) and (c,d) surface coverage ($\eta$) for TPM particles adsorbing at the toluene/water interface from various bulk concentrations ($c_0$). Panels (a,c): pendant drop ($r_d \approx 1~\text{mm}$) ; panels (b,d): microtensiometer drop ($r_d \approx 80~\mu\text{m}$). Experimental data (hollow markers), diffusion–RSA model (solid colored and dashed black lines), and Ward–Tordai model (grey lines). The inset (lower‐left in panel b) shows the drop‐area ratio, $A/A_0$, versus time for each concentration, where $A$ is the instantaneous drop area and $A_0$ is its initial value. The volume fraction is proportional to the concentration $c_b$ shown in the legend. For the RSA model, the maximum surface coverage $\Gamma_m$ determined experimentally corresponds to the RSA jamming limit ($\theta=0.547$).}
\label{fig:data_merged}
\end{figure}

Experimental results are presented in Figure~\ref{fig:data_merged} and compared to two theoretical models. 
The first is a diffusion-limited model based on the Ward–Tordai framework (Section~\ref{Part_WTTheo}) but adapted for particles~\cite{hua2018a}. This model assumes that while partitioning between the bulk and interface can be described by an isotherm, the resulting interfacial tension is solely a function of the particle surface excess via the wetting equation of state (Eq.~\ref{eq:wetting_eos}). This approach decouples the surface pressure from the complexities of the adsorption isotherm, a key distinction from the classical surfactant model. The second is our unified Diffusion–Random Sequential Adsorption (RSA) framework (Section~\ref{Part_RSATheo}), which accounts for irreversible attachment and dynamic blocking effects as the interface becomes progressively crowded.

The Diffusion–RSA model (solid colored and dashed black lines) provides excellent agreement with the experimental data across all concentrations, timescales, and droplet sizes. It captures both the initial adsorption dynamics and the subsequent slowdown as the interface approaches jamming. The Ward–Tordai model performs adequately at early time and relatively low surface coverage ($\theta < 0.5$), where it correctly predicts the drop in interfacial tension. This agreement reflects the fact that, for a pristine interface, adsorption is limited primarily by diffusion from the bulk, a regime common to both models. 

However, the Ward–Tordai model consistently overestimates the adsorbed amount at longer times and higher surface coverages. This discrepancy stems from fundamental physical limitations: the model assumes reversible adsorption, which is unrealistic for particles whose attachment is effectively irreversible. As a result, it fails to capture the transition to kinetic control observed in our experiments and discussed in the next section. In contrast, the Diffusion–RSA model incorporates the key physical ingredients: irreversibility and coverage-dependent steric blocking. Thus, it accurately captures the full adsorption dynamics across all timescales.

A discrepancy is observed between the Diffusion–RSA model and the data for the high-concentration microtensiometer experiments (Figure~\ref{fig:data_merged} b,d). The model predicts a slightly faster adsorption rate than is observed experimentally. This can be attributed to an experimental artifact: the droplet's surface area increases during the experiment, particularly at high particle concentrations where the interfacial tension drops significantly (see inset in Figure~\ref{fig:data_merged}b). This dilation of the interface means the actual surface concentration of particles is lower than what would be expected for a constant area, thus slowing the apparent adsorption dynamics. This issue has been noted in the literature, and some groups now use feedback-controlled systems to maintain a constant surface area during such experiments \cite{iasella2022a}. In contrast, our pendant drop experiments do not suffer from this issue, as the surface area remains constant throughout the measurement, leading to good agreement between the model and the data across all concentrations.

\subsection{Quantification of the Intrinsic Interfacial Kinetics}

The unified Diffusion-RSA model contains the intrinsic adsorption rate constant, $k_a$, as a key kinetic parameter. For each experiment, $k_a$ was determined by fitting the model to the data. The diffusion coefficient $D$ was computed via the Stokes-Einstein relation, and $\Gamma_m$ estimated based on particle geometry and the RSA jamming limit ($\theta = 0.547$).

\begin{figure}[tbhp]
\centering
\includegraphics[width=0.7\linewidth]{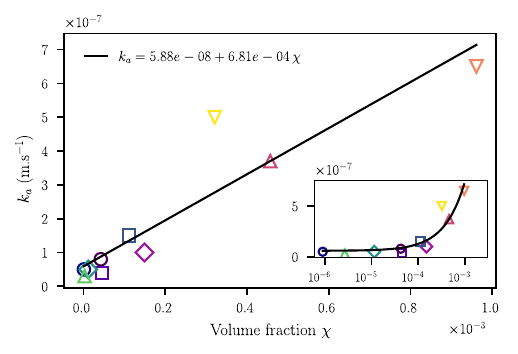}
\caption{Adsorption rate constant $k_a$ vs particle volume fraction $\chi$ from microtensiometer (viridis) and pendant drop (plasma) data. Different particle coatings (TPM-F108 and TPM-Tween). The linear trend $k_a = k_0 + \alpha \chi$ (solid black line) is a global fit.}
\label{fig:kavsc0_combined}
\end{figure}

Figure~\ref{fig:kavsc0_combined} plots the fitted values of $k_a$ versus the particle volume fraction $\chi$ (proportional to $c_b$). A linear dependence, $k_a = k_0 + \alpha \chi$, emerges, robustly consistent across both microtensiometry and pendant drop geometries. This collapse confirms that $k_a$ is an intrinsic, geometry-independent material parameter. The intrinsic adsorption rate constant, $k_a$, extracted from our model fits, is on the order of $10^{-7}$~m/s. 

To place this value in context, we note that for molecular surfactants, experimentally determined values of $k_a$ span several orders of magnitude (from $10^{-8}$ to $10^{-3}$~m/s), reflecting a strong dependence on molecular structure and solution conditions \cite{eastoe2000a, fainerman2006a, he2015a, ferri1999a, bleys1985a, moghimikheirabadi2019a}. Quantifying an equivalent intrinsic rate constant for nanoparticle systems has proven significantly more challenging. 

The clear linear dependence of $k_a$ on the bulk concentration implies that the adsorption flux, which is proportional to $k_a c_b$ in the kinetic regime, contains a quadratic term ($\propto c_b^2$). Such second-order kinetics signify a departure from the simple mechanism of independent, non-interacting particles attaching to the interface. This quadratic dependence points towards a more complex, cooperative adsorption process, analogous to behaviors observed in certain protein and surfactant systems where multimer adsorption or surface-induced aggregation leads to non-linear kinetics \cite{elimelech1998, riechers2016}. Potential physical causes for this cooperativity include attractive interactions between particles in the subsurface that facilitate the formation and subsequent adsorption of dimers or larger clusters. Furthermore, attractive forces between adsorbing particles and those already present at the interface could also play a critical role in lowering the energy barrier for attachment. This latter hypothesis is qualitatively supported by direct confocal imaging of the interface, which reveals the presence of particle aggregates rather than a hexagonally packed monolayer (see Figure~\ref{fig:blocking_with_images}). These results suggest a nuanced picture of the adsorption process: while the RSA framework correctly describes the long-range geometric blocking, the fundamental probability of a particle's attachment, encapsulated in $k_a$, is itself dynamically modulated by significant, concentration-dependent inter-particle interactions.

\section{Rate-Controlling Mechanisms: A Thiele Modulus Analysis}

Having established a validated model for the adsorption process, 
we now seek to quantitatively dissect the evolving balance between bulk diffusion and interfacial kinetics. To achieve this, we employ the Thiele modulus, $\phi$, that compares the characteristic timescale of diffusion, $\tau_D$, to that of the kinetic process, $\tau_k$. For our system, it is defined as:
\begin{equation}
\phi = \frac{\tau_D}{\tau_k}
\end{equation}
In this framework, a large Thiele modulus ($\phi \gg 1$) signifies that the kinetic processes at the interface are intrinsically much faster than the rate at which particles are supplied by diffusion; the system is therefore diffusion-limited. Conversely, a small modulus ($\phi \ll 1$) indicates that diffusive transport is rapid compared to the attachment process, and the system is kinetically-limited. The crossover between these two fundamental regimes occurs at $\phi \approx 1$.

\subsection{Expression of the Thiele Modulus for Irreversible Adsorption}

A pivotal insight for particulate systems is that the kinetic barrier to adsorption is not static. While we have determined an \textit{intrinsic} rate constant, $k_a$, the \textit{effective} rate of adsorption is continuously throttled as the interface becomes crowded. This effective rate necessitates a dynamic formulation of the Thiele modulus that evolves with the normalized surface coverage, $\theta$.

We adapt the theoretical framework for transport to spherical interfaces, originally developed for molecular surfactants by Alvarez et al. \cite{alvarez2010b}, to define the characteristic timescales for our particle system. The diffusion timescale, $\tau_D$, depends on the droplet radius, $r_d$, and on two key length scales: $h_p$, the thickness of a hypothetical bulk layer containing a quantity of particles equivalent to a full monolayer, and $h_s$, the thickness of the spherical depletion shell from which particles are drawn:
\begin{align}
h_p &= \frac{\Gamma_m}{c_b} \\
h_s &= r_d \left[ \left(1 + \frac{3h_p}{r_d} \right)^{1/3} - 1 \right]
\end{align}
The characteristic diffusion time, which elegantly incorporates the spherical geometry, is then given by:
\begin{equation}
\tau_D = \frac{\sqrt{h_s^3 h_p}}{D}
\end{equation}

To obtain an analytical expression for the kinetic timescale $\tau_k$, we require an analytical expression for the surface concentration $\Gamma(t)$. In the regime of high surface coverage, where steric effects dominate, we adopt the empirical expression proposed by Bizmark et al. \cite{bizmark2014a} for the blocking function:
\begin{equation}
B(\theta) \approx 2.32 \, (1 - \theta)^3
\end{equation}
Substituting this into the rate equation~\eqref{eq:adsorption_rate_rsa_model}
and integrating yields the following analytical expression for the surface concentration as a function of time:
\begin{equation}
\Gamma(t) = \Gamma_m \left[ 1 - \sqrt{1 - \left( \frac{1}{4.6 \ k_\text{a,eff} \ c_b \ t / \Gamma_m \ + \ 1} \right)} \right]
\end{equation}
This expression enables direct estimation of the kinetic timescale $\tau_k$, that is inversely proportional to the effective adsorption rate.

Critically, for an RSA process, this rate depends on the available interfacial area, which is captured by the blocking function, $B(\theta)$. This function, which decays from $B(0) \approx 1$ towards zero as $\theta$ approaches the jamming limit, dynamically slows the adsorption. The kinetic timescale is therefore a function of coverage:
\begin{equation}
\tau_k(\theta) = \frac{\Gamma_m}{4.6 k_a c_b B(\theta)}
\end{equation}
Combining these definitions yields the final expression for our dynamic Thiele modulus, $\phi(\eta)$:
\begin{equation}
\phi(\theta) = \frac{\tau_D}{\tau_k(\theta)} = \frac{4.6 k_a B(\theta) \sqrt{h_s^3 h_p}}{D}
\end{equation}
This expression encapsulates the essential physics of the process: a system can begin in a diffusion-limited state, but as the interface fills and $B(\theta)$ diminishes, $\phi(\theta)$ will inevitably decrease, driving the system towards kinetic control.

\subsection{Evolution of Rate Control with Interfacial Coverage}


\begin{figure}[tbhp]
\makebox[\linewidth][c]{%
  \includegraphics[width=0.7\linewidth]{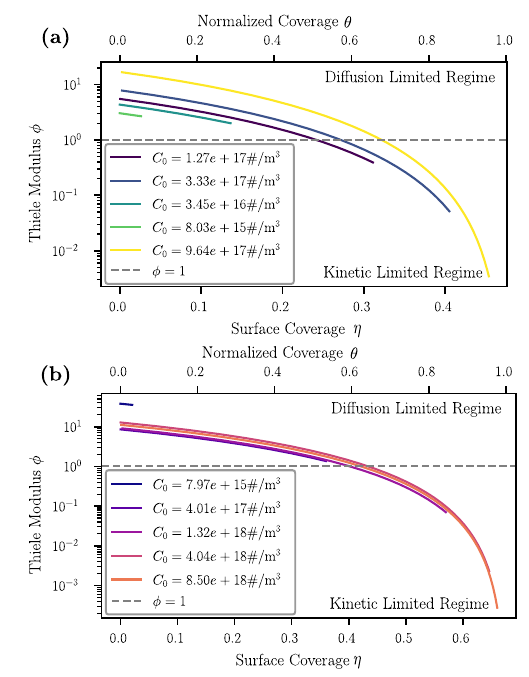}%
}
\captionsetup{width=1\linewidth}
\caption{The Dynamic Thiele Modulus as a Function of Surface Coverage. The evolution of $\phi(\eta)$ is plotted for (a) the 80~µm microtensiometer drop and (b) the 1~mm pendant drop at various bulk concentrations (see legend). The region above the dashed line ($\phi > 1$) corresponds to the diffusion-limited regime, while the region below ($\phi < 1$) corresponds to the kinetically-limited regime. The plots vividly illustrate the transition from diffusion to kinetic control during the course of adsorption. } 
\label{fig:thiele}
\end{figure}

Figure~\ref{fig:thiele} presents the calculated evolution of $\phi(\theta)$ as a function of surface coverage for our experimental systems. This analysis provides a direct, quantitative visualization of the transition in the rate-limiting mechanism. For the large, 1~mm pendant drops (Figure~\ref{fig:thiele}b), the adsorption process begins with $\phi$ values significantly greater than 1, confirming that the initial stages are diffusion-limited. As coverage increases, all trajectories show a decrease in $\phi$, eventually crossing the critical $\phi = 1$ boundary and entering the kinetically-limited regime, where interfacial blocking becomes the primary bottleneck.

The initial values of $\phi$ for the 80~µm microtensiometer drops (Figure~\ref{fig:thiele}a) are substantially lower, often starting near the $\phi = 1$ threshold. This is a direct consequence of the smaller drop radius, $r_d$, which leads to a much shorter diffusion timescale $\tau_D$. The system is therefore more sensitive to the finite rate of interfacial kinetics from the very beginning of the experiment. 

It is worth addressing a common heuristic used in the literature after analyzing here the transition with the Thiele modulus. Several reports, particularly for irreversible adsorption systems, identify the transition from diffusion-limited to kinetically-limited regimes as the point where the adsorption data deviates from a short-time $t^{1/2}$ dependence, a hallmark of pure diffusion to an interface~\cite{bizmark2014a, tian2018a}. While this deviation does indicate that diffusion is no longer the sole rate-limiting step, our analysis reveals a more nuanced picture. The Thiele modulus calculations (Fig.~\ref{fig:thiele}) demonstrate that the system can remain predominantly diffusion-limited ($\phi > 1$) long after the initial $t^{1/2}$ behavior has ceased. This is because the transition is gradual, and significant interfacial blocking is required to shift the balance fully to kinetic control. Our framework thus provides a more rigorous, quantitative criterion for identifying the rate-limiting mechanism, moving beyond the simpler, and potentially misleading, $t^{1/2}$ deviation heuristic.

\subsection{A Universal Map for Adsorption Regime Transitions}

To generalize these findings, the transition between regimes can be projected onto a universal map. This is achieved by recasting the criterion $\phi = 1$ in terms of fundamental dimensionless groups that combine the intrinsic properties of the system with the controllable experimental parameters. Following the approach of Alvarez et al. \cite{alvarez2010b}, we begin by defining a characteristic length scale, $R_{DK}$, which quantifies the intrinsic balance between diffusive transport and interfacial reaction kinetics:
\begin{equation}
R_{DK} = \frac{D}{4.6 k_a}
\end{equation}
 We then define a primary dimensionless group, $q$, which relates this intrinsic length scale to the experimental conditions encapsulated by $h_p$:
\begin{equation}
q = \left( \frac{R_{DK}}{h_p} \right)^{1/3} 
\end{equation}
The condition $\phi = 1$ can be solved for the critical system size, $R_{\text{crit}}$, at which the transition occurs for a given value of $q$. This yields a universal relationship between the dimensionless radius, $R_{\text{crit}}/R_{DK}$, and the dimensionless concentration parameter, $q$:
\begin{equation}\label{eq-map}
\frac{R_{\text{crit}}}{R_{DK}} = \frac{q + \sqrt{(4 - q^2)/3}}{2(1 - q^2)}
\end{equation}

\begin{figure}[htbp]
  \centering
  \includegraphics[width=0.8\linewidth]{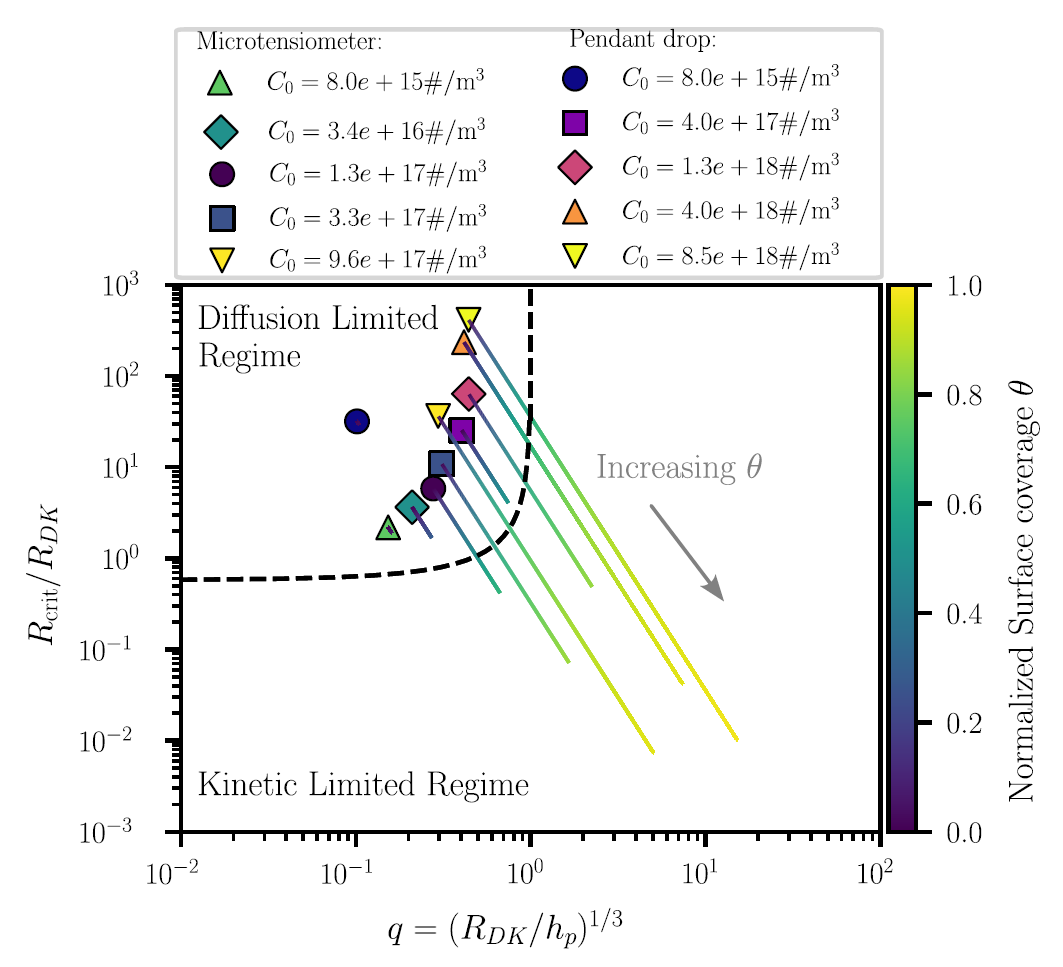}
  \caption{Universal Map of Adsorption Regimes. The transition between diffusion-limited and kinetic-limited regimes is plotted in the dimensionless plane of ($q$, $R_{\text{crit}}/R_{DK}$). The theoretical boundary ($\phi = 1$) derived from Equation~\eqref{eq-map} is shown as the dashed line. Colored trajectories show the evolution of different experimental systems (TPM-F108 and TPM-Tween20) during adsorption. All systems begin in the diffusion-limited regime and move towards the kinetic boundary as surface coverage, $\theta$ (indicated by the color gradient), increases.
  }
  \label{fig:alvarez_transition_map}
\end{figure}

This equation describes a universal boundary in the ($q$, $R_{\text{crit}}/R_{DK}$) plane, separating all possible experimental conditions into either diffusion-limited or kinetically-limited regimes. Figure~\ref{fig:alvarez_transition_map} plots the experimental trajectories on this universal map. Each trajectory begins at a point determined by its initial conditions and evolves as the effective kinetic rate is slowed by interfacial blocking (i.e., as the effective $R_{DK}$ increases with coverage $\theta$).
The colored lines on the map correspond to the simulations of the experiments (Figure~\ref{fig:data_merged}). These lines highlight that, particularly for low-concentration systems, only a small fraction of the full surface coverage is reached within experimental time limits: for low $c_b$, the surface coverage remains well below the maximum threshold throughout the accessible time window.

This universal map demonstrates the advantage of employing multiscale experimental platforms. The microtensiometer experiments, characterized by a smaller droplet radius $R_{\text{crit}}$, are located significantly lower on the vertical axis ($R_{\text{crit}}/R_{DK}$) than their pendant drop counterparts. This vertical displacement positions them closer to the theoretical boundary of kinetic control (the dashed line).

However, the map also reveals a crucial experimental challenge: every single trajectory, including those for the smallest droplets, originates to the left of the $\phi=1$ boundary. This demonstrates that even under the conditions of microtensiometry, the initial adsorption process remains governed by the rate of particle diffusion from the bulk. 

The map therefore serves as a predictive tool, indicating that to design an experiment that is truly kinetically limited from its onset, even smaller systems would be required. Based on our analysis, probing the intrinsic kinetics of these nanoparticles without the confounding influence of diffusion would likely necessitate the use of droplets with radii in the sub-10~$\mu$m range. This finding underscores the profound difficulty in decoupling transport from reaction in colloidal science and highlights the utility of this universal framework in guiding the design of next experiments.

\section{Conclusion}

This work presented a comprehensive analysis of the adsorption dynamics of irreversible colloidal TPM particles at a toluene/water interface, integrating experiments at two different length scales with a unified theoretical framework. The key to successfully capturing the experimental dynamics was the development of a model that couples bulk diffusion with a Random Sequential Adsorption (RSA)-based kinetic boundary condition. Unlike classical diffusion-limited frameworks such as the Ward--Tordai model, this approach correctly accounts for both the particles' irreversible attachment and the crucial role of coverage-dependent steric blocking, which causes a kinetic slowdown at higher surface coverages that classical models cannot predict.

Furthermore, the evolution of the rate-limiting step was quantitatively measured using a dynamic Thiele modulus, $\phi(\theta)$, demonstrating that systems often begin in a diffusion-limited state ($\phi > 1$) but inevitably transition to being kinetically-limited ($\phi < 1$) as the interface fills. The analysis also provided new physical insights into the attachment process itself, revealing that the intrinsic adsorption rate constant, $k_a$, depends linearly on the bulk concentration, $c_b$. This implies the presence of quadratic adsorption kinetics, suggesting that the mechanism is more complex than independent particle events and likely involves cooperative particle-particle interactions in the subsurface or at the interface. These findings collectively advance our understanding of colloidal assembly dynamics and offer a predictive framework for designing experiments and controlling particle-laden interfaces across a range of applications.

The work presented here highlight the necessity of moving beyond classical surfactant models to frameworks that explicitly account for irreversibility and realistic interfacial kinetics, such as RSA blocking. This improved understanding can aid in the rational design of systems involving particle-laden interfaces, from stable Pickering emulsions to the controlled assembly of functional nanoparticle films. Future work could explore the applicability of alternative blocking functions to see which are best suited for different particle systems. The model could also be extended to include specific interfacial interactions, such as the complexation of particles with surfactants, to see how these affect the adsorption dynamics. Such studies would help build a more comprehensive picture, allowing for a better comparison of the adsorption mechanisms of nanoparticles, large macromolecules like proteins, and traditional surfactants, and could help elucidate the transition from reversible to irreversible adsorption.

\section*{Acknowledgements}

This work was supported by the National Science Foundation (NSF) grant CBET 2224412 (ISS: Microgravity enabled studies of particle adsorption dynamics at fluid interfaces).  This project has received funding from the European Union’s Horizon research and innovation programme under the Marie Skłodowska-Curie grant agreement No 101104426 (METABOLISM).

\bibliographystyle{abbrvnat} 
\bibliography{Biblio} 

\begin{thebibliography}{52}
\providecommand{\natexlab}[1]{#1}
\providecommand{\url}[1]{\texttt{#1}}
\expandafter\ifx\csname urlstyle\endcsname\relax
  \providecommand{\doi}[1]{doi: #1}\else
  \providecommand{\doi}{doi: \begingroup \urlstyle{rm}\Url}\fi

\bibitem[Adamczyk(2000)]{adamczyk2000a}
Z.~Adamczyk.
\newblock Kinetics of diffusion-controlled adsorption of colloid particles and proteins.
\newblock \emph{Journal of Colloid and Interface Science}, 229\penalty0 (2):\penalty0 477--489, 2000.
\newblock ISSN 0021-9797.
\newblock \doi{https://doi.org/10.1006/jcis.2000.6993}.

\bibitem[Adamczyk et~al.(2005)Adamczyk, Jaszczolt, Michna, Siwek, Szyk-Warszynska, and Zembala]{adamczyk2005a}
Z.~Adamczyk, K.~Jaszczolt, A.~Michna, B.~Siwek, L.~Szyk-Warszynska, and M.~Zembala.
\newblock Irreversible adsorption of particles on heterogeneous surfaces.
\newblock \emph{Adv Colloid Interface Sci}, 118\penalty0 (1-3):\penalty0 25–42, 2005.
\newblock \doi{10.1016/j.cis.2005.03.003.}

\bibitem[Alvarez et~al.(2010{\natexlab{a}})Alvarez, Walker, and Anna]{alvarez2010a}
N.~Alvarez, L.~Walker, and S.~Anna.
\newblock A microtensiometer to probe the effect of radius of curvature on surfactant transport to a spherical interface.
\newblock \emph{Langmuir}, 26\penalty0 (16):\penalty0 13310–13319, 2010{\natexlab{a}}.
\newblock \doi{10.1021/la101870m.}

\bibitem[Alvarez et~al.(2010{\natexlab{b}})Alvarez, Walker, and Anna]{alvarez2010b}
N.~Alvarez, L.~Walker, and S.~Anna.
\newblock Diffusion-limited adsorption to a spherical geometry: the impact of curvature and competitive time scales.
\newblock \emph{Phys. Rev. E Stat. Nonlin. Soft Matter Phys}, 82\penalty0 (1 Pt 1):\penalty0 011604, 2010{\natexlab{b}}.
\newblock \doi{10.1103/PhysRevE.82.011604.}

\bibitem[Aussillous and Quere(2001)]{aussillous2001a}
P.~Aussillous and D.~Quere.
\newblock Liquid marbles.
\newblock \emph{Nature}, 411\penalty0 (6840):\penalty0 924–927, 2001.
\newblock \doi{10.1038/35082026.}

\bibitem[Berg(2009)]{berg2009a}
J.~Berg.
\newblock \emph{An Introduction to Interfaces and Colloids: The Bridge to Nanoscience}.
\newblock World Scientific Publishing Co. Pte. Ltd, 2009.
\newblock \doi{10.1142/7579.}

\bibitem[Binks(2002)]{binks2002a}
B.~Binks.
\newblock Particles as surfactants—similarities and differences.
\newblock \emph{Curr. Opin. Colloid Interface Sci}, 7\penalty0 (1-2):\penalty0 21–41, 2002.
\newblock \doi{10.1016/s1359-0294(02)00008-0.}

\bibitem[Bizmark et~al.(2014)Bizmark, Ioannidis, and Henneke]{bizmark2014a}
N.~Bizmark, M.~Ioannidis, and D.~Henneke.
\newblock Irreversible adsorption-driven assembly of nanoparticles at fluid interfaces revealed by a dynamic surface tension probe.
\newblock \emph{Langmuir}, 30\penalty0 (3):\penalty0 710–717, 2014.
\newblock \doi{10.1021/la404357j.}

\bibitem[Bleys and Joos(1985)]{bleys1985a}
G.~Bleys and P.~Joos.
\newblock Adsorption kinetics of bolaform surfactants at the air/water interface.
\newblock \emph{J. Phys. Chem}, 89\penalty0 (6):\penalty0 1027–1032, 1985.

\bibitem[Cates and Clegg(2008)]{cates2008a}
M.~E. Cates and P.~S. Clegg.
\newblock Bijels: a new class of soft materials.
\newblock \emph{Soft Matter}, 4:\penalty0 2132--2138, 2008.
\newblock \doi{10.1039/B807312K}.

\bibitem[Chai et~al.(2020)Chai, Hasnain, Bahl, Wong, Li, Geissler, Kim, Jiang, Gu, Li, Lei, Helms, Russell, and Ashby]{chai2020}
Y.~Chai, J.~Hasnain, K.~Bahl, M.~Wong, D.~Li, P.~Geissler, P.~Y. Kim, Y.~Jiang, P.~Gu, S.~Li, D.~Lei, B.~A. Helms, T.~P. Russell, and P.~D. Ashby.
\newblock {Direct observation of nanoparticle-surfactant assembly and jamming at the water-oil interface}.
\newblock \emph{Science Advances}, 6\penalty0 (48):\penalty0 eabb8675, Nov 2020.
\newblock \doi{10.1126/sciadv.abb8675}.

\bibitem[Chen et~al.(2014)Chen, Zhou, Bing, Zhang, Li, Ren, and Qu]{chen2014a}
Z.~Chen, L.~Zhou, W.~Bing, Z.~Zhang, Z.~Li, J.~Ren, and X.~Qu.
\newblock Light controlled reversible inversion of nanophosphor-stabilized pickering emulsions for biphasic enantioselective biocatalysis.
\newblock \emph{J. Am. Chem. Soc}, 136\penalty0 (20):\penalty0 7498–7504, 2014.
\newblock \doi{10.1021/ja503123m.}

\bibitem[Cui et~al.(2017)Cui, Miesch, Kosif, Nie, Kim, Kim, Emrick, and Russell]{cui2017a}
M.~Cui, C.~Miesch, I.~Kosif, H.~Nie, P.~Kim, H.~Kim, T.~Emrick, and T.~Russell.
\newblock Transition in dynamics as nanoparticles jam at the liquid/liquid interface.
\newblock \emph{Nano Lett}, 17\penalty0 (11):\penalty0 6855–6862, 2017.
\newblock \doi{10.1021/acs.nanolett.7b03159.}

\bibitem[Du et~al.(2010)Du, Glogowski, Emrick, Russell, and Dinsmore]{du2010a}
K.~Du, E.~Glogowski, T.~Emrick, T.~Russell, and A.~Dinsmore.
\newblock Adsorption energy of nano- and microparticles at liquid-liquid interfaces.
\newblock \emph{Langmuir}, 26\penalty0 (15):\penalty0 12518–12522, 2010.
\newblock \doi{10.1021/la100497h.}

\bibitem[Eastoe and Dalton(2000)]{eastoe2000a}
J.~Eastoe and J.~Dalton.
\newblock Dynamic surface tension and adsorption mechanisms of surfactants at the air–water interface.
\newblock \emph{Adv. Colloid Interface Sci}, 85\penalty0 (2):\penalty0 103–144, 2000.
\newblock \doi{10.1016/S0001-8686(99)00017-2.}

\bibitem[Elimelech et~al.(1998)Elimelech, Gregory, Jia, and Williams]{elimelech1998}
M.~Elimelech, J.~Gregory, X.~Jia, and R.~A. Williams.
\newblock Particle deposition and aggregation: measurement, modelling and simulation.
\newblock 1998.

\bibitem[Fainerman et~al.(2006)Fainerman, Miller, Ferri, Watzke, Leser, and Michel]{fainerman2006a}
V.~Fainerman, R.~Miller, J.~Ferri, H.~Watzke, M.~Leser, and M.~Michel.
\newblock Reversibility and irreversibility of adsorption of surfactants and proteins at liquid interfaces.
\newblock \emph{Adv Colloid Interface Sci}, 123-126:\penalty0 163–171, 2006.
\newblock \doi{10.1016/j.cis.2006.05.023.}

\bibitem[Feder(1980)]{feder1980}
J.~Feder.
\newblock Random sequential adsorption.
\newblock \emph{J. Theor. Biol.}, 63\penalty0 (1):\penalty0 51--89, 1980.
\newblock \doi{10.1016/0021-9797(80)90358-X}.

\bibitem[Ferri and Stebe(1999)]{ferri1999a}
J.~Ferri and K.~Stebe.
\newblock A structure-property study of the dynamic surface tension of three acetylenic diol surfactants.
\newblock \emph{Colloids Surf., A}, 156\penalty0 (1–3):\penalty0 567–577, 1999.
\newblock \doi{10.1016/S0927-7757(99)00121-1.}

\bibitem[Fink et~al.(2024)Fink, Kim, Han, Kim, To, Bang, Park, Russell, Helms, and Ashby]{fink2024afm}
Z.~Fink, P.~Y. Kim, J.~Han, S.-Y. Kim, N.~To, S.-M. Bang, S.-H. Park, T.~P. Russell, B.~A. Helms, and P.~D. Ashby.
\newblock {Repairable and Reconfigurable Structured Liquid Circuits}.
\newblock \emph{Advanced Functional Materials}, 34\penalty0 (15):\penalty0 2311406, 2024.

\bibitem[Fu et~al.(2024)Fu, Zhao, Chen, Zhang, and Chai]{fu2024a}
Y.~Fu, S.~Zhao, W.~Chen, Q.~Zhang, and Y.~Chai.
\newblock Self-assembly of nanoparticles with stimulated responses at liquid interfaces.
\newblock \emph{Nano Today}, 54:\penalty0 102073, 2024.
\newblock \doi{10.1016/j.nantod.2023.102073.}

\bibitem[Garbin et~al.(2012)Garbin, Crocker, and Stebe]{garbin2012a}
V.~Garbin, J.~Crocker, and K.~Stebe.
\newblock Forced desorption of nanoparticles from an oil-water interface.
\newblock \emph{Langmuir}, 28\penalty0 (3):\penalty0 1663–1667, 2012.
\newblock \doi{10.1021/la202954c.}

\bibitem[Guzman et~al.(2022)Guzman, Martinez-Pedrero, Calero, Maestro, Ortega, and Rubio]{guzman2022a}
E.~Guzman, F.~Martinez-Pedrero, C.~Calero, A.~Maestro, F.~Ortega, and R.~Rubio.
\newblock A broad perspective to particle-laden fluid interfaces systems: from chemically homogeneous particles to active colloids.
\newblock \emph{Adv Colloid Interface Sci}, 302:\penalty0 102620, 2022.
\newblock \doi{10.1016/j.cis.2022.102620.}

\bibitem[He et~al.(2015)He, Yazhgur, Salonen, and Langevin]{he2015a}
Y.~He, P.~Yazhgur, A.~Salonen, and D.~Langevin.
\newblock Adsorption-desorption kinetics of surfactants at liquid surfaces.
\newblock \emph{Adv Colloid Interface Sci}, 222:\penalty0 377–384, 2015.
\newblock \doi{10.1016/j.cis.2014.09.002.}

\bibitem[Hill and Eastoe(2017)]{hill2017a}
C.~Hill and J.~Eastoe.
\newblock Foams: From nature to industry.
\newblock \emph{Advances in Colloid and Interface Science}, 247:\penalty0 496--513, 2017.
\newblock \doi{10.1016/j.cis.2017.05.013}.

\bibitem[Hua et~al.(2016)Hua, Bevan, and Frechette]{hua2016a}
X.~Hua, M.~Bevan, and J.~Frechette.
\newblock Reversible partitioning of nanoparticles at an oil-water interface.
\newblock \emph{Langmuir}, 32\penalty0 (44):\penalty0 11341–11352, 2016.
\newblock \doi{10.1021/acs.langmuir.6b02255.}

\bibitem[Hua et~al.(2018)Hua, Frechette, and Bevan]{hua2018a}
X.~Hua, J.~Frechette, and M.~Bevan.
\newblock Nanoparticle adsorption dynamics at fluid interfaces.
\newblock \emph{Soft Matter}, 14\penalty0 (19):\penalty0 3818–3828,, 2018.
\newblock \doi{10.1039/c8sm00273h.}

\bibitem[Iasella et~al.(2022)Iasella, Barman, Ciutara, Huang, Davidson, and Zasadzinski]{iasella2022a}
S.~Iasella, S.~Barman, C.~Ciutara, B.~Huang, M.~Davidson, and J.~Zasadzinski.
\newblock Microtensiometer for confocal microscopy visualization of dynamic interfaces.
\newblock \emph{J Vis Exp}, 187:\penalty0 64110, 2022.
\newblock \doi{10.3791/64110.}

\bibitem[Kaz et~al.(2012)Kaz, McGorty, Mani, Brenner, and Manoharan]{kaz2012a}
D.~Kaz, R.~McGorty, M.~Mani, M.~Brenner, and V.~Manoharan.
\newblock Physical ageing of the contact line on colloidal particles at liquid interfaces.
\newblock \emph{Nat. Mater}, 11\penalty0 (2):\penalty0 138–142, 2012.
\newblock \doi{10.1038/nmat3190.}

\bibitem[Lin et~al.(1990)Lin, McKeigue, and Maldarelli]{lin1990a}
S.~Lin, K.~McKeigue, and C.~Maldarelli.
\newblock Diffusion‐controlled surfactant adsorption studied by pendant drop digitization.
\newblock \emph{AIChE journal}, 36\penalty0 (12):\penalty0 1785–1795, 1990.

\bibitem[Manga et~al.(2016)Manga, Hunter, Cayre, York, Reichert, Anna, Walker, Williams, and Biggs]{manga2016a}
M.~Manga, T.~Hunter, O.~Cayre, D.~York, M.~Reichert, S.~Anna, L.~Walker, R.~Williams, and S.~Biggs.
\newblock Measurements of submicron particle adsorption and particle film elasticity at oil-water interfaces.
\newblock \emph{Langmuir}, 32\penalty0 (17):\penalty0 4125–4133, 2016.
\newblock \doi{10.1021/acs.langmuir.5b04586.}

\bibitem[Moghimikheirabadi et~al.(2019)Moghimikheirabadi, Fischer, Kroger, and Sagis]{moghimikheirabadi2019a}
A.~Moghimikheirabadi, P.~Fischer, M.~Kroger, and L.~Sagis.
\newblock Relaxation behavior and nonlinear surface rheology of peo-ppo-peo triblock copolymers at the air-water interface.
\newblock \emph{Langmuir}, 35\penalty0 (44):\penalty0 14388–14396, 2019.
\newblock \doi{10.1021/acs.langmuir.9b02540.}

\bibitem[Neibloom et~al.(2020)Neibloom, Bevan, and Frechette]{neibloom2020a}
D.~Neibloom, M.~Bevan, and J.~Frechette.
\newblock Surfactant-stabilized spontaneous 3-(trimethoxysilyl) propyl methacrylate nanoemulsions.
\newblock \emph{Langmuir}, 36\penalty0 (1):\penalty0 284–292, 2020.
\newblock \doi{10.1021/acs.langmuir.9b03412.}

\bibitem[Pan et~al.(2023)Pan, Gao, Ge, Gao, Huang, Kang, Luo, Zhang, Fan, Zhu, and Wang]{pan2023a}
Y.~Pan, S.~Gao, C.~Ge, Q.~Gao, S.~Huang, Y.~Kang, G.~Luo, Z.~Zhang, L.~Fan, Y.~Zhu, and A.~Wang.
\newblock Removing microplastics from aquatic environments: A critical review.
\newblock \emph{Environ Sci Ecotechnol}, 13:\penalty0 100222, 2023.
\newblock \doi{10.1016/j.ese.2022.100222.}

\bibitem[Park and Lee(2014)]{park2014a}
B.~Park and D.~Lee.
\newblock Particles at fluid–fluid interfaces: From single-particle behavior to hierarchical assembly of materials.
\newblock \emph{MRS Bull}, 39\penalty0 (12):\penalty0 1089–1098, 2014.
\newblock \doi{10.1557/mrs.2014.253.}

\bibitem[Peito et~al.(2022)Peito, Peixoto, Ferreira-Faria, Margarida~Martins, Margarida~Ribeiro, Veiga, Marto, and Claudia Paiva-Santos]{peito2022a}
S.~Peito, D.~Peixoto, I.~Ferreira-Faria, A.~Margarida~Martins, H.~Margarida~Ribeiro, F.~Veiga, J.~Marto, and A.~Claudia Paiva-Santos.
\newblock Nano- and microparticle-stabilized pickering emulsions designed for topical therapeutics and cosmetic applications.
\newblock \emph{Int J Pharm}, 615:\penalty0 121455, 2022.
\newblock \doi{10.1016/j.ijpharm.2022.121455.}

\bibitem[Prosser and Franses(2001)]{prosser2001a}
A.~Prosser and E.~Franses.
\newblock Adsorption and surface tension of ionic surfactants at the air–water interface: review and evaluation of equilibrium models.
\newblock \emph{Colloids Surf. Physicochem. Eng. Aspects}, 178\penalty0 (1-3):\penalty0 1--40, 2001.
\newblock \doi{10.1016/S0927-7757(00)00552-5}.

\bibitem[Riechers et~al.(2016)Riechers, Souza, Krüger, and Windbergs]{riechers2016}
B.~Riechers, A.~M. Souza, M.~Krüger, and M.~Windbergs.
\newblock ph-responsive surfactants for controlled drug delivery applications.
\newblock \emph{Int J Pharm}, 509:\penalty0 1--14, 2016.
\newblock \doi{10.1016/j.ijpharm.2016.05.020}.

\bibitem[Schwenke et~al.(2014)Schwenke, Isa, and Del~Gado]{schwenke2014a}
K.~Schwenke, L.~Isa, and E.~Del~Gado.
\newblock Assembly of nanoparticles at liquid interfaces: crowding and ordering.
\newblock \emph{Langmuir}, 30\penalty0 (11):\penalty0 3069–3074, 2014.
\newblock \doi{10.1021/la404254n.}

\bibitem[Shi and Russell(2018)]{cui2019sa}
S.~Shi and T.~P. Russell.
\newblock Nanoparticle assembly at liquid–liquid interfaces: From the nanoscale to mesoscale.
\newblock \emph{Advanced Materials}, 30\penalty0 (44):\penalty0 1800714, 2018.
\newblock \doi{https://doi.org/10.1002/adma.201800714}.

\bibitem[Shin et~al.(2019)Shin, Huang, Neibloom, Bevan, and Frechette]{shin2019a}
D.~Shin, T.~Huang, D.~Neibloom, M.~Bevan, and J.~Frechette.
\newblock Multifunctional liquid marble compound lenses.
\newblock \emph{ACS Appl Mater Interfaces}, 11\penalty0 (37):\penalty0 34478–34486, 2019.
\newblock \doi{10.1021/acsami.9b12738.}

\bibitem[Shipway et~al.(2000)Shipway, Katz, and Willner]{shipway2000a}
A.~Shipway, E.~Katz, and I.~Willner.
\newblock Nanoparticle arrays on surfaces for electronic, optical, and sensor applications.
\newblock \emph{Chemphyschem}, 1\penalty0 (1):\penalty0 18–52, 2000.

\bibitem[Slavchov and Ivanov(2017)]{slavchov2017a}
R.~Slavchov and I.~Ivanov.
\newblock Adsorption parameters and phase behaviour of non-ionic surfactants at liquid interfaces.
\newblock \emph{Soft Matter}, 13\penalty0 (46):\penalty0 8829–8848, 2017.
\newblock \doi{10.1039/c7sm01370a.}

\bibitem[Smirnov et~al.(2015)Smirnov, Peljo, Scanlon, and Girault]{smirnov2015a}
E.~Smirnov, P.~Peljo, M.~Scanlon, and H.~Girault.
\newblock Interfacial redox catalysis on gold nanofilms at soft interfaces.
\newblock \emph{ACS Nano}, 9\penalty0 (6):\penalty0 6565–6575, 2015.
\newblock \doi{10.1021/acsnano.5b02547.}

\bibitem[Talbot et~al.(2000)Talbot, Tarjus, Tassel, and Viot]{talbot2000a}
J.~Talbot, G.~Tarjus, P.~Tassel, and P.~Viot.
\newblock From car parking to protein adsorption: an overview of sequential adsorption processes.
\newblock \emph{Colloids Surf. A Physicochem. Eng. Asp}, 165\penalty0 (1-3):\penalty0 287–324, 2000.
\newblock \doi{10.1016/S0927-7757(99)00409-4.}

\bibitem[Thompson et~al.(2015)Thompson, Williams, and Armes]{thompson2015a}
K.~Thompson, M.~Williams, and S.~Armes.
\newblock Colloidosomes: synthesis, properties and applications.
\newblock \emph{J Colloid Interface Sci}, 447:\penalty0 217–228, 2015.
\newblock \doi{10.1016/j.jcis.2014.11.058.}

\bibitem[Tian et~al.(2018)Tian, Feng, Cho, Datta, and Prud'homme]{tian2018a}
C.~Tian, J.~Feng, H.~Cho, S.~Datta, and R.~Prud'homme.
\newblock Adsorption and denaturation of structured polymeric nanoparticles at an interface.
\newblock \emph{Nano Lett}, 18\penalty0 (8):\penalty0 4854–4860, 2018.
\newblock \doi{10.1021/acs.nanolett.8b01434.}

\bibitem[Vethaak and Legler(2021)]{vethaak2021a}
A.~Vethaak and J.~Legler.
\newblock Microplastics and human health.
\newblock \emph{Science}, 371\penalty0 (6530):\penalty0 672–674, 2021.
\newblock \doi{10.1126/science.abe5041.}

\bibitem[Vialetto et~al.(2020)Vialetto, Rudiuk, Morel, and Baigl]{vialetto2020a}
J.~Vialetto, S.~Rudiuk, M.~Morel, and D.~Baigl.
\newblock From bulk crystallization of inorganic nanoparticles at the air/water interface: tunable organization and intense structural colors.
\newblock \emph{Nanoscale}, 12\penalty0 (11):\penalty0 6279–6284, 2020.
\newblock \doi{10.1039/c9nr10965j.}

\bibitem[Vialetto et~al.(2021)Vialetto, Rudiuk, Morel, and Baigl]{vialetto2021a}
J.~Vialetto, S.~Rudiuk, M.~Morel, and D.~Baigl.
\newblock Photothermally reconfigurable colloidal crystals at a fluid interface, a generic approach for optically tunable lattice properties.
\newblock \emph{J. Am. Chem. Soc}, 143\penalty0 (30):\penalty0 11535–11543, 2021.
\newblock \doi{10.1021/jacs.1c04220.}

\bibitem[Wang et~al.(2016)Wang, McGorty, Kaz, and Manoharan]{wang2016a}
A.~Wang, R.~McGorty, D.~Kaz, and V.~Manoharan.
\newblock Contact-line pinning controls how quickly colloidal particles equilibrate with liquid interfaces.
\newblock \emph{Soft Matter}, 12\penalty0 (43):\penalty0 8958–8967, 2016.

\bibitem[Ward and Tordai(1946)]{ward1946a}
A.~Ward and L.~Tordai.
\newblock Time‐dependence of boundary tensions of solutions i. the role of diffusion in time‐effects.
\newblock \emph{J. Chem. Phys}, 14\penalty0 (7):\penalty0 453–461, 1946.
\newblock \doi{10.1063/1.1724167.}

\end{thebibliography}

\end{document}